\documentclass[%
article,
twocolumn,
superscriptaddress,
nofootinbib,
amsmath,amssymb,
aps,
]{revtex4-1}

\usepackage{graphicx}
\usepackage{dcolumn}
\usepackage{bm}
\usepackage{hyperref}
\usepackage{float}%

\usepackage{amsmath}
\usepackage{MnSymbol}

\usepackage{makecell}

\begin{document}
	
	
	\title{Machine Learning and Evolutionary Prediction of Superhard B-C-N Compounds}
	
	\author{Wei-Chih Chen}
	\affiliation{Department of Physics, University of Alabama at Birmingham, Birmingham, Alabama 35294, USA}
	
	\author{Joanna N. Schmidt}
	\affiliation{Department of Physics, University of Alabama at Birmingham, Birmingham, Alabama 35294, USA}
	
	\author{Da Yan}
	\affiliation{Department of Computer Science, University of Alabama at Birmingham, Birmingham, Alabama 35294, USA}
	
	\author{Yogesh K. Vohra}
	\affiliation{Department of Physics, University of Alabama at Birmingham, Birmingham, Alabama 35294, USA}
	
	\author{Cheng-Chien Chen$^*$}
	\affiliation{Department of Physics, University of Alabama at Birmingham, Birmingham, Alabama 35294, USA}
	\footnotetext{Corresponding author email: chencc@uab.edu}
	
	\date{\today}
	
	\begin{abstract}
		We build random forests models to predict elastic properties and mechanical hardness of a compound, using only its chemical formula as input. The model training uses over 10,000 target compounds and 60 features based on stoichiometric attributes, elemental properties, orbital occupations, and ionic bonding levels.
		Using the models, we construct triangular graphs for B-C-N compounds to map out their bulk and shear moduli, as well as hardness values. The graphs indicate that a 1:1 B-N ratio can lead to various superhard compositions. We also validate the machine learning results by evolutionary structure prediction and density functional theory. Our study shows that BC$_{10}$N, B$_4$C$_5$N$_3$, and B$_2$C$_3$N exhibit dynamically stable phases with hardness values $>40$GPa, which are potentially new superhard materials that could be synthesized by low-temperature plasma methods.
	\end{abstract}
	
	\maketitle
	
	\section{Introduction}
	
	Superhard materials exhibit a Vickers hardness $H \ge 40$ GPa, and they have extensive applications such as abrasives, cutting tools, and protective coatings~\cite{friedrich2011synthesis, zhao2016recent, yeung2016ultraincompressible, kvashnin2019computational, le2019high}.
	Diamond is the hardest material (with $H \sim 100$ GPa), but its applications are limited by size and cost~\cite{Haines_review}. 
	It is also not suitable for oxidizing conditions or high-speed machining of ferrous alloys because of chemical reactions with iron-group elements~\cite{Haines_review,Zhao_review}.
	One promising class of superhard materials involve light elements B, C, N, and O~\cite{Kurakevych2009}.
	These elements can form multiple short covalent bonds, which make the crystal structure difficult to break apart.
	One notable example is cubic boron nitride (c-BN), which has a reported hardness between $H\sim 50- 70$ GPa~\cite{ZHANG2014607}.
	Other examples are boron carbides, ranging from weakly boron-incorporated diamond structures like BC$_5$~\cite{PhysRevLett.102.015506,li2010superhard,baker2018computational} to boron-rich B$_{12}$ icosahedron-intercalated structures like B$_{50}$C$_2$~\cite{uemura2016structure, Baker2020,Chakrabarty_2020}.
	For ternary compounds, several superhard B-C-N compositions have been reported~\cite{liu2011synthesis, PhysRevB.93.144107}, such as BC$_2$N~\cite{SOLOZHENKO20012228,BC2N_BC4N_2002} and BC$_4$N~\cite{BC2N_BC4N_2002}.
	Other superhard B-C-O~\cite{wang2016novel,liu2017superhard}, B-N-O~\cite{li2015superhard, bhat2015high}, and C-N-O~\cite{steele2017ternary} compounds also have been studied.
	However, due to the huge phase space of possible element combinations, 
	it remains challenging to explore new superhard ternary materials.
	
	First-principles simulations based on density functional theory have played important roles in predicting new superhard compounds.
	However, {\it ab initio} methods are still computationally expensive and size-limited.
	On the other hand, data-driven approaches have proven to be powerful and efficient in exploring new materials~\cite{schmidt2019recent, zhou2019big, himanen2019data,chibani2020machine,saal2020machine,cai2020machine} -- thanks to recent advance in computing hardwares, development in machine learning algorithms, and availability of online materials database.
	For example, Meredig {\it et al.}~\cite{PhysRevB.89.094104} have constructed a machine learning model to {\it screen over 1.6 million ternary compositions} and predicted 4,500 novel, potentially stable ternary materials.
	Therefore, data-driven machine learning approaches are promising for large-scale materials design and discovery.
	
	In principle, a machine learning framework can be implemented with different material features or descriptors for a wide range of target properties. Two popular properties to predict are bulk and shear moduli~\cite{furmanchuk2016predictive, deJong2016, Isayev2017, evans2017predicting,Mansouri2018, Avery2019}, which are also correlated with the material hardness.
	For example, de Jong {\it et al.}~\cite{deJong2016} developed a technique based on gradient boosting and used features like the volume per atom and cohesive energy.
	Mansouri {\it et al.}~\cite{Mansouri2018} used support vector machines and combined elemental and structural properties as descriptors, where the cohesive energy was also identified as a crucial feature.
	These machine learning studies typically can achieve high prediction accuracy with only a few thousands of training data points.
	However, using cohesive energy, volume, melting point, crystal symmetry and so on as features may be less ideal, as obtaining these information for new compounds would require additional measurements or calculations.
	
	In this paper, we develop random forests models to predict material mechanical properties, by using only features that can be derived directly from the chemical formula. The resulting machine learning models thereby can achieve large-scale prediction of new superhard and ultra incompressible materials for extreme environment applications.
	We also employ evolutionary structure prediction and density functional theory calculations to further validate the machine learning results.
	In particular, we propose three new superhard compositions -- BC$_{10}$N, B$_4$C$_5$N$_3$, and B$_2$C$_3$N -- and fully characterize their structural, phonon, and electronic properties. These new superhard compounds are all dynamically stable with relatively low formation energy, so they can potentially be synthesized by low-temperature plasma methods, {\it without the need of} high-temperature high-pressure conditions. It is noted that our newly suggested compound BC$_{10}$N has a computed hardness value $\sim 87$ GPa; once synthesized, the compound would become the second hardest material. Our computational flowchart is summarized in Fig. 1 and discussed in detail in the Methods section. 
	
	\begin{figure*}[!th]
		\begin{center}
			\includegraphics[width=\textwidth]{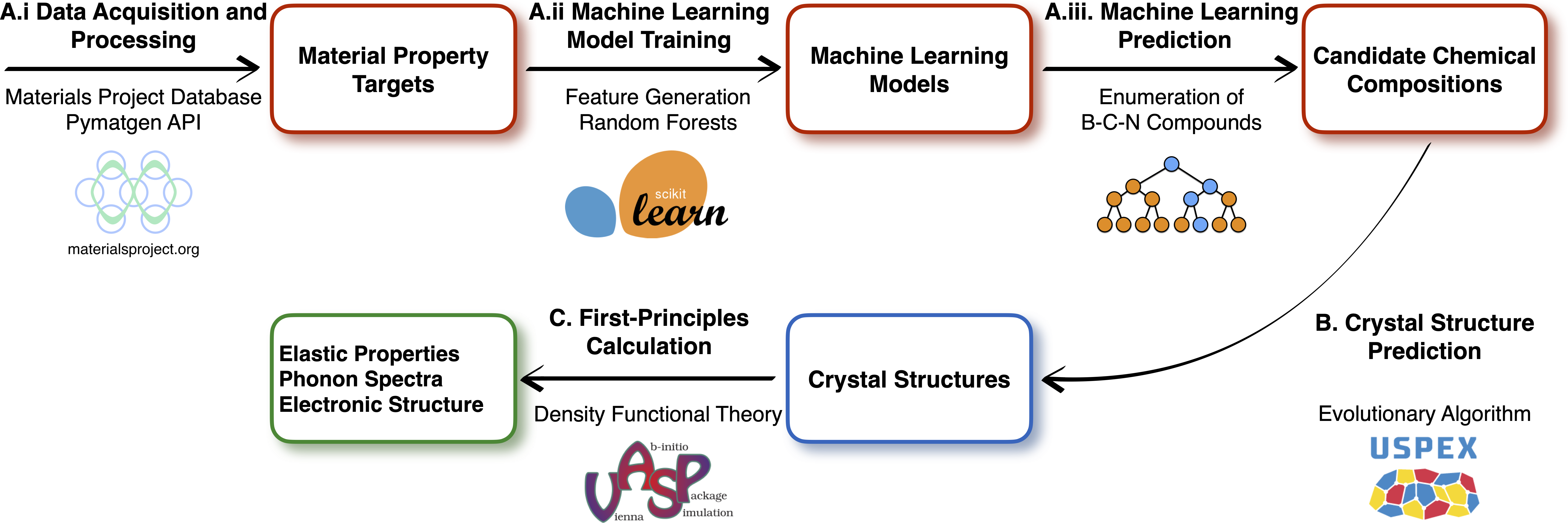}
			\caption{\label{fig:flowchart}
				Computational flowchart of data-driven discovery of new superhard materials in this study: {\bf A.i} Data acquisition and processing using the Materials Project~\cite{MP} database and its application programming interface (API) \textsc{Pymatgen}~\cite{PyMatGen}; {\bf A.ii} Machine learning model training with handcrafted features and regression algorithms implemented in the \textsc{scikit-learn} library~\cite{scikit-learn}; {\bf A.iii} Random forests prediction of chemical compositions for candidate superhard materials. {\bf B.} Crystal structure prediction of a given chemical formula using evolutionary algorithms implemented in the USPEX program~\cite{oganov2006crystal, glass2006uspex, lyakhov2013new}. {\bf C.} First-principles validation of the machine learning results with density functional theory calculations using the VASP software~\cite{kresse1996efficiency,kresse1996efficient}.
			}
		\end{center}
	\end{figure*}
	
	\section{Computational Methods}

	\subsection{Machine Learning Model}
	
	{\it Data acquisition --}
	There exist several online computational materials databases, such as AFLOW~\cite{aflow}, Materials Project~\cite{MP}, NOMAD Encyclopedia~\cite{nomad}, and the Open Quantum Materials Database (OQMD)~\cite{OQMD}. Here we use the Materials Project~\cite{MP}, which provides open access to various computed properties of known and predicted crystalline compounds. The corresponding Python Materials Genomics (\textsc{Pymatgen}) library~\cite{PyMatGen} is utilized to extract the target properties of bulk modulus ($K$) and shear modulus ($G$).
	The Materials Project~\cite{MP} database also contains high-pressure phases and artificial crystal structures, which can exhibit extreme values of bulk and shear moduli. Therefore, we exclude those extreme outliers and focus on 10,421 selected compounds with $K$ and $G$ values both in the ranges of $0-550$ GPa.
	
	{\it Feature generation --} To build a supervised learning model using only chemical composition as input, we generate features (or descriptors) based on a compound's chemical formula. By following Ref.~\onlinecite{ward2016general}, we consider features related to stoichiometric attributes, elemental properties, orbital occupations, and ionic levels.
	Part of the features can be generated with the Python library \textsc{Matminer}~\cite{matminer}. 
	We {\emph{do not}} consider structural or electronic features like crystal symmetry, volume, melting point, band gap, etc. While including these additional features could improve the model performance, these information is {\textit {a priori}} unknown for new compounds. To expedite materials discovery, we thereby do not include features that require additional first-principles calculations.
	
	First, the stoichiometric features are computed using the $L^p$ norm $||x||_p = (\sum_i |x_i|^p)^{1/p}$, where $x_i$ is element $i$'s atomic fraction. These attributes capture the changes in atomic fraction, independent of the actual elements. As an example, the $p=2$ norm of Fe$_2$O$_3$ is $||x||_2 = \left((\frac{2}{5})^{2} + (\frac{3}{5})^2\right)^{1/2} \simeq 0.721$~\cite{ward2016general}. Here we consider 3 stoichiometric features, including the $p=0$ norm (i.e. the number of chemical components), and the $p=2, 3$ norms. The $p=1$ norm is equal to unity regardless of the chemical composition, so it is not considered. In addition, we do not find an apparent model improvement with more higher order norms ($p>3$), so they are not included.
	
	Second, the elemental features are computed using the minimum, maximum, and range for properties of each element present, as well as the values of fraction-weighted mean $\bar f = \sum_i x_i f_i$ and average deviation $\hat f = \sum_i | f_i - \bar f|$. Here, $f_i$ is the property of element $i$, and $x_i$ is the atomic fraction. We consider the following 10 properties: atomic number, atomic mass, element column number, row number, atomic radius, electronegativity, and the numbers of valence electrons in $s$, $p$, $d$, and $f$ orbitals, respectively. Therefore, there are 50 elemental-property features (= 5 values $\times$ 10 properties).
	Using again Fe$_2$O$_3$ as an example~\cite{ward2016general}, for the ``atomic number" property, $\bar f = \frac{2}{5}(26) + \frac{3}{5}(8) = 15.2$ and $\hat f = \frac{2}{5}|26 - 15.2| + \frac{3}{5}|8-15.2|= 8.64$. 
	
	Third, 4 orbital-occupation features are computed using the fraction-weighted average of the number of valance electrons respectively in $s$, $p$, $d$, and $f$ orbitals, divided by the fraction-weighted average of the total number of valance electrons. For example, Fe$_2$O$_3$'s $p$-orbital occupation feature is $F_p = \frac{2/5\times(0) + 3/5\times(4)}{2/5\times(8)+3/5\times(6)} \simeq 0.353$~\cite{ward2016general}.
	
	Finally, 3 features are based on ionic levels. The first is a Boolean number denoting whether it is possible to form a neutral ionic compound, by assuming that each element takes exactly one of its common charge states. The other two features are based on the ``ionic character'' of a chemical bond: $I(\chi_i, \chi_j)= 1- \exp( - (\chi_i-\chi_j)^2/4)$, where $\chi_i$ and $\chi_j$ are electronegativities for elements $i$ and $j$, respectively. In Pauling scale, fluorine has the highest electronegativity value of $\chi_F= 3.98$, and francium has the lowest electronegativity value of $\chi_{Fr} = 0.70$. The two features we consider are respectively the maximum ionic character $I$ between any two elements in a compound, and the mean ionic character $\bar I = \sum_{i,j} x_i x_j \chi_i \chi_j$.
	
	In total, 60 features are created. To simplify the training task, we do not consider additional feature engineering such as degree-2 polynomials, which otherwise could lead to thousands of new features and cause overfitting. The chemical compositions and their target properties of bulk and shear moduli for the 10,421 compounds considered here are written as a Python dictionary object saved in a \textsc{json} file. The features for all compounds are provided as a \textsc{csv} file accordingly. Both files are downloadable from the Electronic Supporting Information (ESI).
	
	{\it Model training, validation, and application --} For regression task, we choose the random forests algorithm~\cite{ho1998, amit1997shape},
	which is a tree-based ensemble method. A random forests model builds multiple decision trees, by taking a random sample with replacement from the training set and a random subset of features to split tree nodes. The results averaged over individual trees serve as the final predictions, which help reduce variance and improve accuracy. However, without restriction on the tree depth, the model can become very deep and cause overfitting. Therefore, we constrain the pre-pruning parameter of tree depth to regularize the models.
	
	The model training is implemented with the \textsc{scikit-learn} library~\cite{scikit-learn}. We use 90$\%$ of our samples as the training and validation set, which is then used to determined the tree depth by 10-fold cross-validation. The remaining 10$\%$ is the test set used for an unbiased evaluation of the final model. We build two separate models to predict the bulk modulus ($K$) and shear modulus ($G$), respectively. We do not train a model for predicting the Vickers hardness ($H$), as the target hardness value is not as widely available as $K$ and $G$. On the other hand, there exist several empirical models for evaluating $H$~\cite{hardnessmodel2003,hardnessmodel2006,hardnessmodel2008,hardnessmodel2011,niu2019simple, mazhnik2019model}, based on physical properties such as bond length, bond strength, electronegativity, and covalent radius. Here we adopt hardness models that require only bulk and shear moduli as inputs~\cite{chen2011modeling, tian2012microscopic}, so that our regression results of $K$ and $G$ can be employed directly to predict $H$. 
	
	After training and evaluation, we apply the models to predict mechanical properties of B-C-N compounds and search for new superhard ternary materials. For candidate compositions identified with superhardness (i.e. $H \ge$ 40 GPa), we then perform crystal structure prediction and first-principles calculations to further validate the machine learning predictions.
	
	\subsection{Crystal Structure Prediction}
	Crystal structure prediction (CSP) concerns finding the stable structure of a compound knowing only its chemical formula~\cite{wang2014perspective,graser2018machine, oganov2019structure}. In principle, this is achieved by locating the minimum of the Gibbs free energy $G_{\textrm{free}} = U + PV - TS$, where $U$ is the total energy, $P$ is the pressure, $V$ is the volume, $T$ is the temperature, and $S$ is the entropy. In practice, the entropy and temperature effects are often neglected, and only the enthalpy $H_{\textrm{free}} = U  + PV$ is minimized. The minima of the potential energy surface correspond to different stable and metastable structures, which could be stabilized under different $P-T$ conditions.
	
	CSP requires an accurate estimation of the system's total energy $U$ (usually from first-principles calculation), and an efficient optimization technique. Here we utilize the highly efficient implementation of evolutionary algorithm in USPEX (Universal Structure Predictor: Evolutionary Xtallography)~\cite{oganov2006crystal, glass2006uspex, lyakhov2013new}. Evolutionary algorithm is a population-based optimization technique using biological evolution concepts such as mutation, recombination, and selection.
	Candidate solutions are individuals in a population,
	which will evolve after applications of the above operators and selection by a fitness function.
	For a given composition, we examine over at least 1,000 structures.
	The first generation of structures are randomly created.
	Subsequent generations are created with $20\%$ from random structures and $80\%$ from heredity, softmutation, and transmutation operators.
	
	\subsection{First-Principles Calculation}
	Our first-principles density functional theory (DFT) calculations are performed with the Vienna Ab initio Simulation Package (VASP)~\cite{kresse1996efficiency,kresse1996efficient}.
	The Monkhorst-Pack sampling scheme~\cite{monkhorst1976special} is used with a $\Gamma$-centered $k$-point mesh of $21\times 21 \times 5$ (resolution = 0.02$\times$2$\pi$/$\text{\normalfont\AA}$) points in the Brillouin zone.
	The convergence criteria of self-consistent and structural relaxation calculations are set to 10$^{-6}$ eV/unit-cell and 10$^{-3}$ eV/$\text{\normalfont\AA}$, respectively.
	We adopt a plane wave energy cutoff of 520 eV, which is sufficient to converge the DFT total energy difference $< 10^{-4}$ eV/atom.
	For each crystal structure, we first fully relax the lattice parameters and atomic positions.
	After structure relaxation, we then compute the corresponding mechanical, electronic, and phonon properties.
	All calculations use projector augmented wave (PAW)~\cite{blochl1994projector, kresse1999ultrasoft} pseudopotentials and the Perdew-Burke-Ernzerhof generalized gradient approximation (GGA) functional~\cite{perdew1996generalized}.
	
	We employ the strain-stress method~\cite{PhysRevB.65.104104} in VASP to compute the elastic constants $C_{ij}$,
	which in turn can determine the bulk and shear moduli using the Vogit-Reuss-Hill formula~\cite{voigt1928lehrbuch, reuss1929berechnung, hill1952elastic}.
	Phonon dispersion spectra are computed using the \textsc{Phonopy} package~\cite{phonopy}.
	Density functional perturbation theory with $2\times 2\times 1$ supercells are adopted to evaluate the second-order force constants. 
	
	\begin{figure*}[!th]
		\begin{center}
			\includegraphics[width=\textwidth]{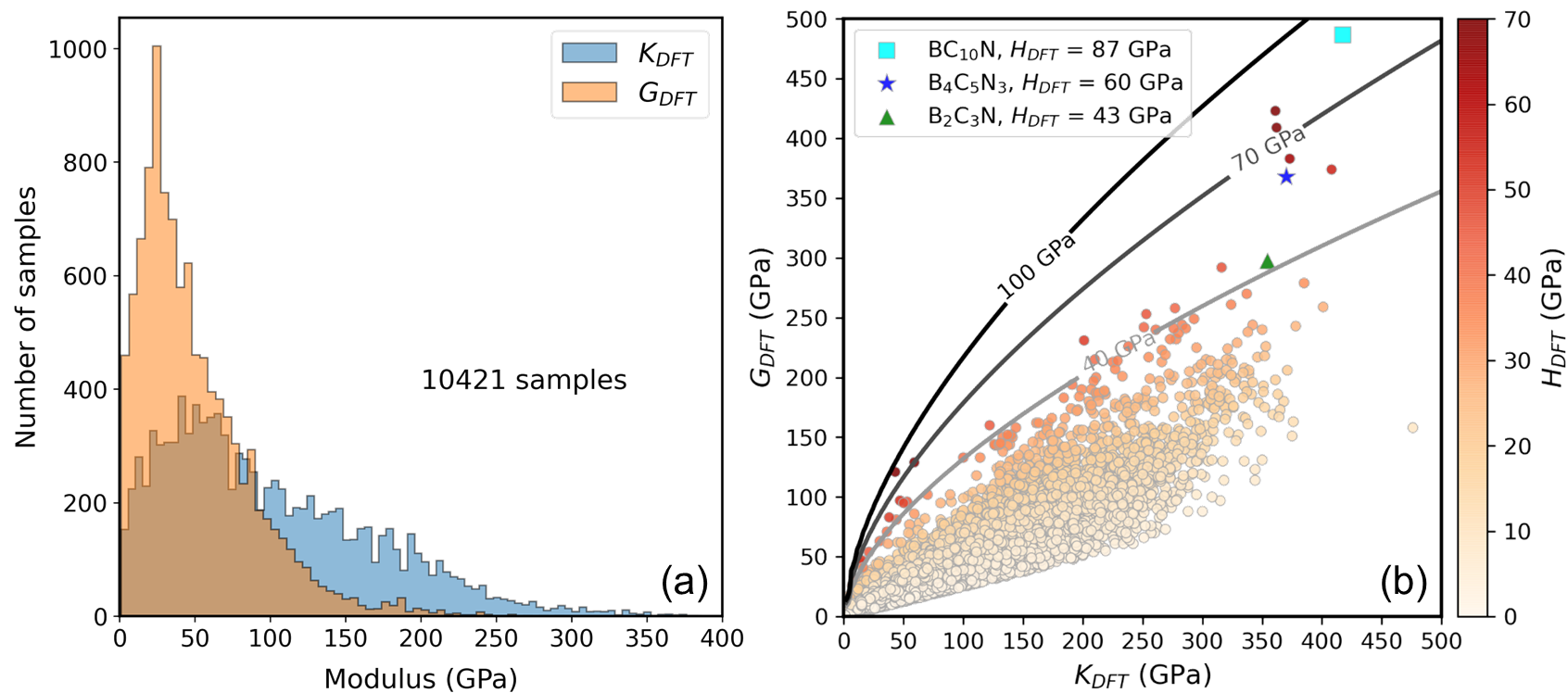}
			\caption{\label{fig:ml}
				(a) Histogram and (b) scatter plot of bulk ($K$) and shear ($G$) moduli for 10,421 samples acquired from the Materials Project~\cite{MP} database based on density functional theory (DFT) calculations. In (b), the false-color intensity represents the Vickers hardness ($H$) computed by Tian's empirical model using $K$ and $G$ as inputs. The solid curves represent the hardness contours using Tian's model~\cite{tian2012microscopic}.
				These contour lines can help quickly locate compounds with superhardness, with the caveat that the model's applicability might be more limited in the low-bulk/low-shear modulus region.
				The three newly proposed superhard compounds BC$_{10}$N, B$_4$C$_5$N$_3$, and B$_2$C$_3$N are highlighted respectively by the $\square$, $\medstar$, and $\triangle$ symbols.
			}
		\end{center}
	\end{figure*}
	
	\section{Results and Discussion}
	
	Figure 2(a) shows histograms of the bulk and shear moduli computed by DFT (denoted respectively as $K_{DFT}$ and $G_{DFT}$) for 10,421 samples acquired from the Materials Project database~\cite{MP}. Here, the DFT modulus values represent the Voigt-Reuss-Hill average moduli~\cite{voigt1928lehrbuch, reuss1929berechnung, hill1952elastic}, and the medians of $K_{DFT}$ and $G_{DFT}$ in Fig. 2(a) are 84 GPa and 40 GPa, respectively. Since the accuracy and applicability of a machine learning model largely depend on the training data, we have set a few criteria to select suitable samples during data acquisition.
	
	First, we have excluded sample materials with a formation energy $\ge 0.2$ eV/atom, as they are thermodynamically unfavorable.
	Second, we have neglected samples whose Voigt and Reuss modulus values differ by more than 50 GPa; this class of samples are typically layered quasi-two-dimensional materials, like graphite or hexagonal boron nitride, which are not the focus of our study.
	Third, we have utilized the Pugh's ratio $k$ $(\equiv G/K)$~\cite{Pugh} to further filter out materials with $k < 0.25$ due to their extremely small hardness, as well as materials with $k > 4.0$, which represents an extreme high-hardness structure (with $H > 200$ GPa) usually computed under high pressure.
	With these selection criteria, there are in total 10,421 samples considered in our machine learning study.
	
	Figure 2(b) shows the scatter plot for the distribution of bulk and shear moduli. The false-color intensity represents the corresponding material hardness ($H$), which is calculated by using Tian's empirical model~\cite{tian2012microscopic}:
	\begin{equation}
		\label{eq:2}
		\begin{aligned}
			H = 0.92k^{1.137}G^{0.708}.
		\end{aligned}
	\end{equation}
	This empirical formula dictates that superhardness requires a large Pugh's ratio $k$ and/or a high shear modulus $G$.
	Using Tian's model, we also plot hardness contour lines in Fig. 2(b). Materials in the contour region between $H= 40 - 100$ GPa are mostly compounds like C, BC$_2$N, c-BN, and M$_x$B$_y$ (M: Be or transition metal). The three newly proposed superhard ternary compounds -- BC$_{10}$N, B$_4$C$_5$N$_3$, and B$_2$C$_3$N -- are highlighted respectively by the $\square$, $\medstar$, and $\triangle$ symbols in Fig. 2(b). These materials will be discussed later in the paper.
	
	\begin{figure*}[!th]
		\begin{center}
			\includegraphics[width=\textwidth]{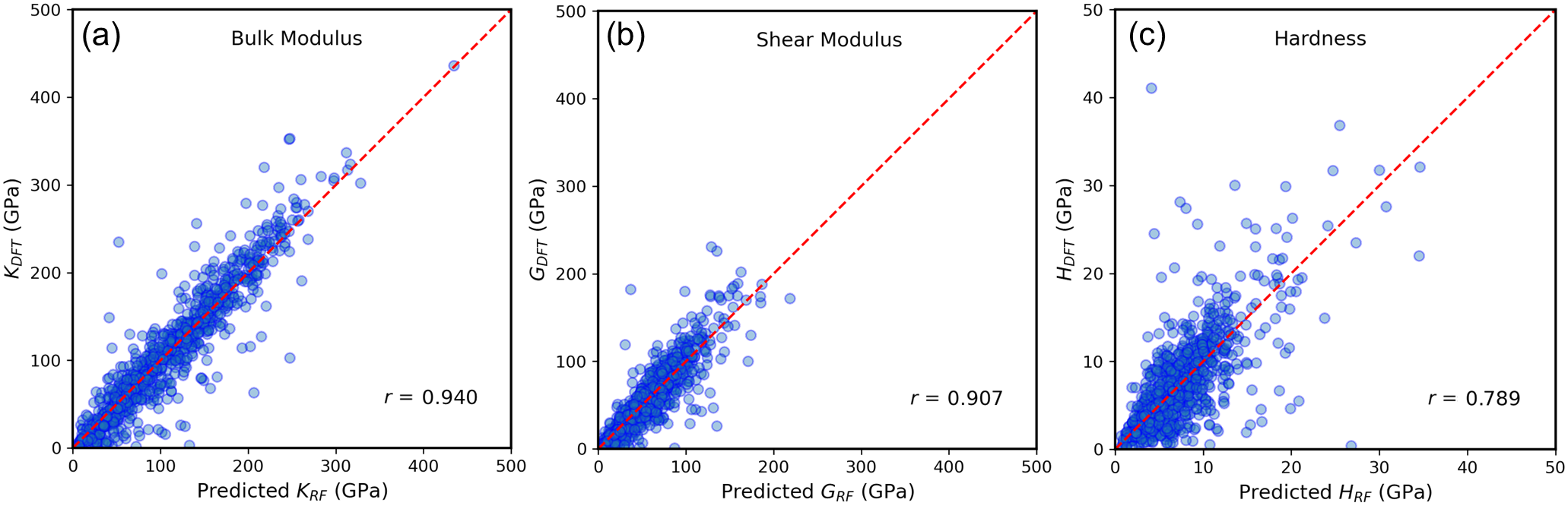}
			\caption{\label{fig:ml}
				Evaluation of random forests (RF) models using the Pearson correlation coefficient ($r$) as a metric, for (a) bulk modulus ($K$), (b) shear modulus ($G$), and (c) hardness $(H)$. The machine learning models are trained to predict respectively $K$ and $G$, and both can achieve $r > 0.9$ when applied to the test set based on density functional theory (DFT) calculations.
				The predicted hardness $H_{RF}$ is obtained by using Tian's empirical formula with $K_{RF}$ and $G_{RF}$ as inputs, which results in an inferior correlation coefficient as anticipated.
			}
		\end{center}
	\end{figure*}
	
	After data acquisition, we split the whole data into the training-validation set (90\%) and the test set (10\%). The training-validation set is used for grid search with 10-fold cross validation to search for a proper tree depth of the random forests models. We find that a maximum depth of 12 layers (with 100 estimators) is reasonable for obtaining a good balance between bias and variance.
	A deeper tree would not improve the model performance.
	After deciding on the maximum tree depth, we also utilize the training-validation set to further refine the machine learning models. 
	For an unbiased evaluation of the final model performance, we use the test set and consider the metric of Pearson correlation coefficient $r$:
	\begin{equation}
		\label{eq:3}
		\begin{aligned}
			r =
			\frac{ \sum_{i=1}^{n}(x_i-\bar{x})(y_i-\bar{y}) }{
				\sqrt{\sum_{i=1}^{n}(x_i-\bar{x})^2}\sqrt{\sum_{i=1}^{n}(y_i-\bar{y})^2}}.
		\end{aligned}
	\end{equation}
	Here, $x_i$ is the machine-learning predicted value for a single entry, and $y_i$ is the corresponding ``actual" value computed by DFT. $\bar{x}$ and $\bar{y}$ represent respectively the mean values of the predicted and the actual (DFT) values in the test set, which contains $n \sim 1,000$ sample points. The $r$ value can range between -1 to 1, and $r=1$ means that the prediction is 100\% accurate.
	
	Figure 3 shows the $r$-value plots using the test set. For bulk and shear moduli [Fig. 3(a) and 3(b)], the data distribution follows closely the $r=1$ dashed line. The $r$ values of $K$ and $G$ are respectively 0.940 and 0907 (, and their coefficients of determination $r^2$ are respectively 0.885 and 0.822). For the hardness $H_{RF}$ in Fig. 3(c), we note that the prediction is {\it not} obtained directly from a machine learning model. Instead, we use the machine learning predicted $K_{RF}$ and $G_{RF}$ with Tian's empirical formula in Eq. (1) to compute $H_{RF}$. Therefore, the data distribution in Fig. 3(c) is more dispersing with a $r$-value $\sim 0.79$, which is slightly inferior as expected.
	We also note that a higher $r$-value (or $r^2$ score) could be achieved by including additional features like volume, crystal symmetry, and cohesive energy, as done in previous studies~\cite{furmanchuk2016predictive, deJong2016, Isayev2017, evans2017predicting,Mansouri2018, Avery2019}. However, here we do not consider features that require additional measurements or calculations, but focus only on features that can be derived directly from the chemical formula, in order to achieve efficient large-scale materials discovery.
	
	Our machine learning models also provide information on feature importance to help reveal features that are more correlated with the bulk or shear moduli. Among the 60 features in our study, the atomic radius and $d$ electron occupation are the most important ones. 
	For example, the average atomic radius of a given compound and the bulk modulus exhibit a negative correlation with $r \sim -0.24$.
	Similarly, the $r$ value between the largest atomic radius and the bulk modulus is $r \sim -0.41$.
	The results indicate that in general, a smaller crystal unit cell will favor a higher bulk modulus.
	This is consistent with the facts that most superhard materials consist of light and small elements like Be, B, C, N, and O, and that diamond has the smallest volume per unit cell among all crystalline materials.
	On the other hand, the $d$ electron occupation rate is positively correlated with the bulk modulus, with a $r$ value $\sim 0.58$.
	This corresponds to the fact that many ultra incompressible materials are transition-metal borides like ReB$_2$ and Os$_2$B$_3$~\cite{Burrage_2020_ReB2,Burrage_2020_Os2B3}.
	
	\begin{figure*}[!th]
		\begin{center}
			\includegraphics[width=\textwidth]{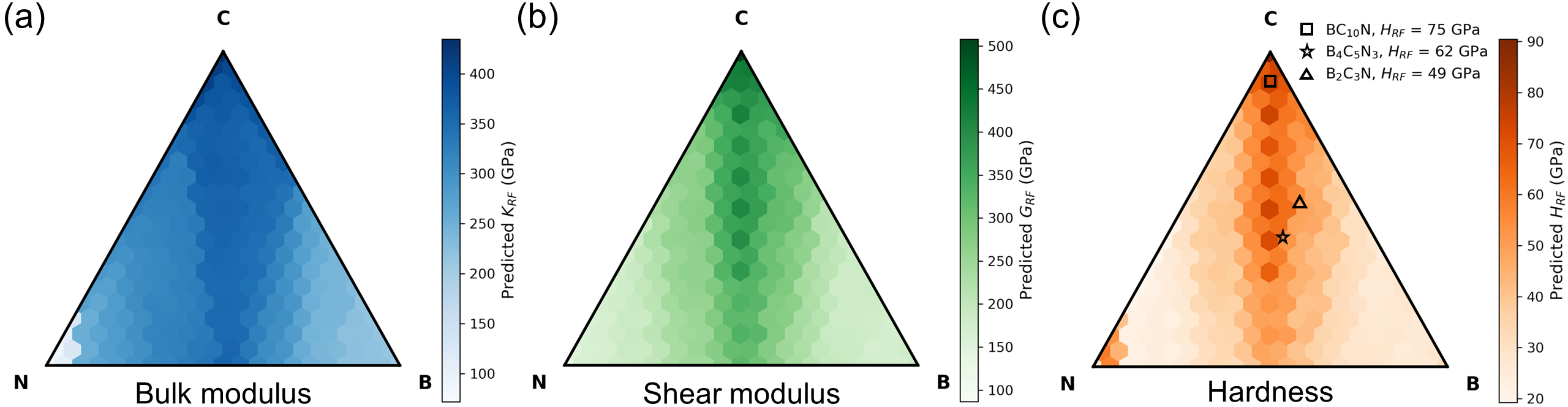}
			\caption{\label{fig:triangles}
				Triangular (or ternary) B-C-N graphs for (a) bulk modulus ($K$), (b) shear modulus ($G$), and (c) hardness $(H)$, predicted by random forests (RF) machine learning models. Panel (c) indicates that a 1:1 B-N composition ratio can lead to various superhard compounds such BC$_2$N ($H_{RF} = 74$ GPa) and BC$_4$N ($H_{RF}=65$ GPa). The hardness of three newly proposed superhard compounds, BC$_{10}$N, B$_4$C$_5$N$_3$, and B$_2$C$_3$N are consistent with the DFT results. The triangular graphs are visualized by using the Python Ternary Plots library~\cite{pythonternary}.
			}
		\end{center}
	\end{figure*}
	
	With the random forest models, we can predict quickly the mechanical properties of a given chemical formula. Here we apply our models to ternary B-C-N compounds by enumerating a series of B$_x$C$_y$N$_z$ compositions, with $x,y,z\in \{1,2,3,...\}$.
	Figure 4 shows the predicted triangular graphs,
	where the corner points correspond to unitary elemental compounds.
	For example, the pure boron phase in Fig. 4(c) is predicted to have $H_{RF}$ $\sim$ 30 GPa, which could be regarded as the hardness for $\alpha$-B, $\beta$-B, $\gamma$-B, or tetragonal B$_{52}$. For the pure carbon phase, the predicted $H_{RF} \sim 90$ GPa could be related to cubic or hexagonal diamond (lonsdaleite).
	One caveat is that the relatively high hardness predicted near the pure nitrogen phase may be unrealistic; this is due to small bulk moduli of nitrogen-dominated compounds, which leads to a large Pugh's ratio $k$ and a high hardness when Tian's model in Eq. (1) is used. If we implement more data selection rules by restricting the $K$ and $G$ values to be $>50$ GPa, then the artifact near the pure nitrogen phase could be avoided. However, this would cause overestimation in the overall mechanical properties.
	
	Figure 4 also shows that B-C-N compositions with a $1:1$ B:N ratio can result in several superhard compounds with hardness $>60$ GPa. For example, the predicted hardness values of BC$_2$N and BC$_4$N by machine learning are 74 and 65 GPa, respectively. These predictions are consistent with previous experimental findings of superhardness in BC$_2$N (76 or 62 GPa)~\cite{SOLOZHENKO20012228,BC2N_BC4N_2002} and BC$_4$N (68 GPa)~\cite{BC2N_BC4N_2002}, which are synthesized under high-pressure and high-temperature conditions. 
	
	Motivated by the machine learning results in Fig. 4, we next employ evolutionary prediction with USPEX to search for potential superhard structures around the region with a B:N ratio $\sim$ 1:1. The calculations are performed under an applied pressure of 15 GPa to help locate stable structures of smaller volumes and larger hardness.
	We first consider 15 trial chemical formulae with even number of valence electrons, including BC$_3$N, BC$_5$N, BC$_6$N, BC$_3$N$_2$, B$_2$C$_3$N, B$_4$C$_5$N$_2$, etc., with a single-formula unit cell.
	However, most of the structures we found are graphite-like structures with $sp^2$ bonding, so they are not superhard.
	On the other hand, we find a diamond-like structure with $sp^3$ bonding for B$_2$C$_3$N [Fig. 5(b)], which exhibits a hardness value $> 40$ GPa and a relatively low formation energy as discussed later.
	
	B$_2$C$_3$N is hexagonal with a superlattice structure along the (111) direction of cubic diamond. By comparing the $1\times 1 \times 2$ supercell of B$_2$C$_3$N (i.e. B$_4$C$_6$N$_2$), we find that such structure is similar to the $1\times 1 \times 2$ BC$_4$N~\cite{BC4N_Luo2008} (i.e. B$_2$C$_8$N$_2$) with 2 carbon atoms replaced by 2 boron atoms, or the structure of $1\times 1 \times 3$ BC$_2$N~\cite{BC2N_Liu2018} (i.e. B$_3$C$_6$N$_3$) with 1 nitrogen replaced by 1 boron. 
	In fact, such atomic replacement is also the case of boron-substituted diamond BC$_5$~\cite{PhysRevB.80.094106} ($H \sim 70$ GPa) in a 12-atom unit cell, with 2 carbon atoms replaced by 2 boron atoms. If one further replaces a carbon by boron in BC$_5$, the resulting B$_2$C$_4$ (BC$_2$)~\cite{Xu2010} structure is also superhard ($H \sim 56$ GPa). 
	Similarly, based on the 12-atom unit cell of diamond, the aforementioned superhard structure of BC$_2$N~\cite{BC2N_Liu2018} (BC$_4$N~\cite{BC4N_Luo2008}) also can be generated by replacing 6 (4) carbons with 3 (2) BN pairs.
	
	Using a 12-atom unit cell with a 1:1 B:N ratio, we first create the structure of BC$_{10}$N [Fig. 5 (a)] by replacing 2 carbon atoms in diamond by 1 pair of BN. Our random forests models predict that BC$_{10}$N has bulk and shear moduli equal to $K_{RF}=379$ GPa and $G_{RF}= 422$ GPa, respectively, which corresponds to a hardness $H_{RF}=75$ GPa by Tian's model.
	Using a similar rule, we also generate a new superhard composition B$_4$C$_5$N$_3$, by replacing 1 carbon with 1 boron in B$_3$C$_6$N$_3$.
	B$_4$C$_5$N$_3$ is predicted to have $K_{RF}=359$ GPa and $G_{RF}=369$ GPa, with $H_{RF}=62$ GPa.
	Other superhard B-C-N compounds also could be generated in a similar way. For example, B$_3$C$_7$N$_2$ and B$_2$C$_9$N could be obtained respectively from B$_3$C$_6$N$_3$ and B$_2$C$_8$N$_2$ by atomic substitution.
	Before we shift the focus to first-principles DFT validation of machine learning results, some comments are in order:
	(i) We have considered boron substitution in a 8-atom unit cell of BC$_2$N (i.e. B$_2$C$_4$N$_2$) to obtain B$_3$C$_3$N$_2$. However, we find that B-C-N compounds in a 8-atom unit cell have higher formation energies $>$ 300 meV/atom, which is consistent with early study on BC$_2$N by Chen {\it et al.}~\cite{PhysRevLett.98.015502}. (ii) We did not consider nitrogen substitution, because such structures tend to be thermodynamically more unstable.
	
	\begin{figure*}[!th]
		\begin{center}
			\includegraphics[width=\textwidth]{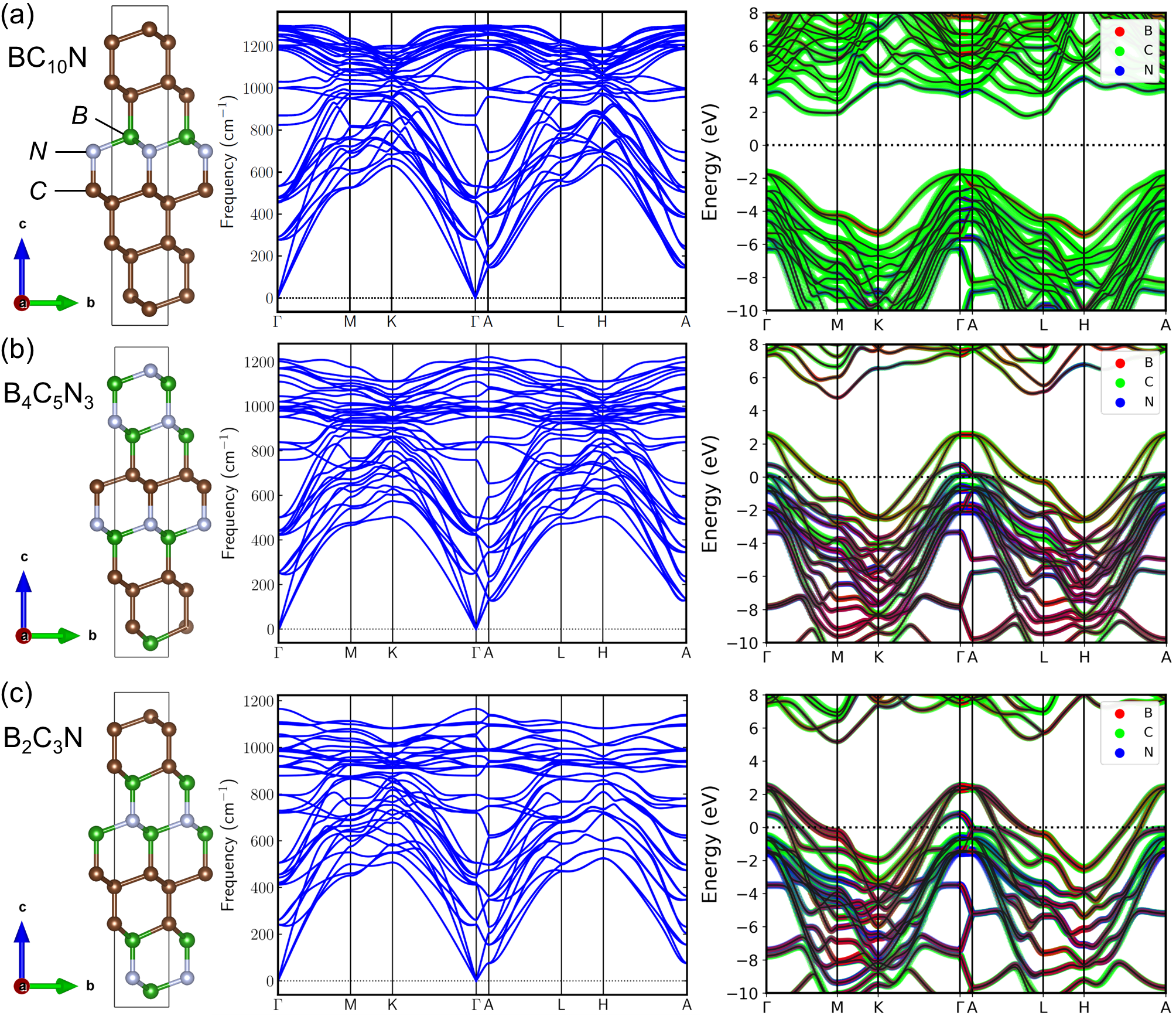}
			\caption{\label{fig:ml}
				Theoretical crystal structures (left panels), phonon dispersion spectra (middle panels), and electronic band structures (right panels) from evolutionary algorithm and density functional theory calculations for (a) BC$_{10}$N, (b) B$_4$C$_5$N$_3$, and (c) B$_2$C$_3$N. BC$_{10}$N is a wide-band-gap insulator, while B$_4$C$_5$N$_3$ and B$_2$C$_3$N are both metals. All three compounds are dynamically stable (i.e. without negative phonon modes). The crystal structures are visualized by the VESTA software~\cite{momma2011vesta}.
			}
		\end{center}
	\end{figure*}
	
	We next discuss DFT calculations of three new superhard B-C-N phases predicted by machine learning: BC$_{10}$N, B$_4$C$_5$N$_3$, and B$_2$C$_3$N.
	The structure of B$_2$C$_3$N [Fig. 5(c)] is discovered by USPEX using a 6-atom unit cell. For BC$_{10}$N [Fig. 5(a)] and B$_4$C$_5$N$_3$ [Fig. 5(b)], they are generated by using the aforementioned rule of atomic substitution in a 12-atom diamond unit cell. We have performed additional USPEX calculations for BC$_{10}$N and B$_4$C$_5$N$_3$, but did not find any lower enthalpy structure.
	The structures in Fig. 5 can be treated as superlattices along the (111) direction of cubic diamond. All structures are trigonal systems with the hexagonal lattice space group $P3m1$ (No. 156), which has 6 independent elastic constants $C_{ij}$: $C_{11}$, $C_{12}$, $C_{13}$, $C_{14}$, $C_{33}$, $C_{44}$, ($C_{66}$ = $(C_{11}-C_{12})/2$). There are 4 necessary and sufficient mechanical stability conditions based on Born's criteria~\cite{PhysRevB.90.224104}: $C_{11} > C_{12}$, $C_{44} > 0$, $C_{13}^2  < \frac{1}{2}C_{33}(C_{11}+C_{12})$, and $C_{14}^2 < \frac{1}{2}C_{44}(C_{11}-C_{12})$.
	The three structures in Fig. 5 all fulfill these criteria, so they are mechanically stable. 
	

	\begingroup
	\setlength{\tabcolsep}{6pt} 
	\renewcommand{\arraystretch}{1.5} 
	\begin{table*}[t]
		\caption{Structural and mechanical properties calculated by density functional theory (DFT), including lattice parameters $a (=b)$ and $c$ ($\text{\normalfont\AA}$) in 12-atom hexagonal unit cell, density $\rho$ (atom/$\text{\normalfont\AA}^3$), bulk modulus $K$ (GPa), shear modulus $G$ (GPa), Young's modulus $E$ (GPa), Pugh's ratio $k$, Poisson's ratio $\nu$, universal elastic anisotropy $A^U$, hardness $H$ (GPa), and formation energy $\Delta$E (meV/atom).}
		\label{tab:table2}
		\begin{tabular}{l|c|c|c|c|c|c|c|c|c|c|c|c} 
			Formula   & $a(=b)$ & $c$ & $\rho$ & $K$ & $G$ & $E$ & $k$ & $\nu$ & $A^U$ & $H_{RF}$ & $H_{DFT}$ & $\Delta$E\\
			\hline
			BC$_{10}$N  & 2.533 & 12.453 & 0.173 & 417  & 487 & 1052 & 1.166 & 0.080 & 0.058 & 75 & 87 & 79.5 \\
			B$_4$C$_5$N$_3$ & 2.558 & 12.722 & 0.166 & 370  & 368 & 829 & 0.995 & 0.125 & 0.221 & 62 & 60 & 141.3 \\
			B$_2$C$_3$N & 2.570 & 12.832 & 0.164 & 354  & 298 & 697 & 0.840 & 0.172 & 0.792 & 49 & 43 & 155.9 \\
			diamond & 2.527 & 12.379 & 0.175 & 432 & 518 & 1110 & 1.199 & 0.072 & 0.044 &90 & 94 & 0
		\end{tabular}
	\end{table*}
	\endgroup

	Figure 5 middle panels show phonon spectra for the corresponding structures in the left panels, and all structures are dynamically stable without negative modes. Their phonon dispersions are fairly similar, due to the similarity in the crystal structures.
	Among them, BC$_{10}$N has the highest phonon frequency above 1250 cm$^{-1}$, while B$_4$C$_5$N$_3$ and B$_2$C$_3$N have slightly lower phonon frequencies near 1200 cm$^{-1}$ at the top of the phonon bands. 
	We note that cubic diamond has the highest phonon frequency above 1300 cm$^{-1}$. The results indicate that phonon bands are softened with increasing B/N content. This phonon softening is consistent with the phenomenon observed in boron-incorporated diamond BC$_5$~\cite{ma11081279}.
	
	The electronic band structures are shown accordingly in Fig. 5 right panels. BC$_{10}$N exhibits a wide band gap $\sim$ 3.5 eV, so it is a superhard insulator. On the other hand, the electron-deficit B$_4$C$_5$N$_3$ and B$_2$C$_3$N are metals, where their valence band maximums are shifted towards the conduction bands. Due to the similarity of phonon and electronic dispersion relations between B$_4$C$_5$N$_3$/B$_2$C$_3$N and BC$_5$~\cite{ma11081279,PhysRevB.80.094106, BC5_SC}, superconductivity may be observed in the predicted B$_4$C$_5$N$_3$ and B$_2$C$_3$N compounds as well. However, discussion related to superconducting properties is beyond the scope of this study. 
	
	The elastic constants $C_{ij}$ computed by DFT for the structures in Fig. 5 can be utilized to derive other mechanical properties, such as the bulk modulus ($K$), shear modulus ($G$), and Young's modulus ($E$). The Vicker's hardness $H$ also can be calculated by Eq.(1) with $K$ and $G$ as input. Details of the calculation results are given in Table I.
	Notably, BC$_{10}$N has supreme mechanical properties with a hardness of 87 GPa, which is comparable to diamond. Once synthesized, BC$_{10}$N would be the second known hardest material.
	
	Table I also indicates that when the B/N content increases, the mechanical strengths like bulk and shear moduli as well as hardness will tend to decrease, which is consistent with the trend of phonon softening. The computed Cauchy pressures ($= C_{12}-C_{44}$)~\cite{PhysRevB.86.134106} of BC$_{10}$N, B$_4$C$_5$N$_3$, and B$_2$C$_3$N are all negative (and equal to -405 GPa, -209 GPa, and -121 GPa, respectively), which suggests their brittle properties and strong covalent bondings. 
	In addition, the Pugh's ratio ($k = G/K$)~\cite{Pugh} is greater than 0.571 for all three ternary compounds, indicative of their brittle properties as well. 
	By increasing the B/N content, other B-C-N compounds like B$_4$C$_5$N$_3$ and B$_2$C$_3$N can become more ductile.
	Table I also demonstrates that the density $\rho$ (atom/$\text{\normalfont\AA}^3$) is positively correlated with mechanical strength, while the poisson's ratio $\nu$ and universal elastic anisotropy $A^U$~\cite{Univ_Ela_Aniso} have negative correlations with elastic moduli and hardness.
	
	Finally, we evalulate the thermodynamic stability of the newly proposed B$_x$C$_y$N$_z$ compounds, by calculating the formation energy $\Delta E$ (also shown in Table I):
	\begin{equation}
		\label{eq:5}
		\begin{aligned}
			\Delta E = \frac{E(B_xC_yN_z)-yE(C) - zE(BN) - (x-z)E(B)}{x+y+z},
		\end{aligned}
	\end{equation}
	which is the difference between the total energy $E$ of B$_x$C$_y$N$_z$ and the atomically weighted reference total energies $E$ of diamond, cubic BN, and $\alpha$-B. The formation energies of the three compounds are positive, which is consistent with previous study on BC$_2$N~\cite{PhysRevLett.98.015502}, suggesting that $\Delta E$ increases due to B-C and C-N bondings.
	Among the three proposed compositions, BC$_{10}$N has the least number of B-C and B-N bonds, and it has the lowest formation energy $<$ 100 meV/atom. Since $\Delta E$ of BC$_{10}$N is smaller than those of BC$_{2}$N and BC$_{4}$N, it is likely that BC$_{10}$N can be synthesized without extreme conditions~\cite{aykol2018thermodynamic}, using e.g. low-temperature plasma methods.

	\section{Conclusion}
	
	We have built random forest models to predict bulk and shear moduli by using target elastic properties in the Materials Project database~\cite{MP}.
	The machine learning models utilize only materials features that can be derived directly from a given chemical formula, so they are suitable for large-scale materials characterization and discovery. We have applied the resulting models to B-C-N compounds to search for new superhard ternary materials.
	The machine-learning predicted ternary graphs indicate that a 1:1 B:N ratio can lead to various promising superhard materials with hardness $>$ 40 GPa.
	We also have utilized evolutionary structure prediction together with first-principles density-functional-theory calculations to further validate the machine learning results.
	We have proposed three new potential superhard ternary compounds -- BC$_{10}$N, B$_4$C$_5$N$_3$, and B$_2$C$_3$N -- and fully characterized their properties. 
	In predicted ternary compounds, BC$_{10}$N is a wide band-gap semiconductor, while B$_4$C$_5$N$_3$ and B$_2$C$_3$N show metallic behavior. 
	Among them, BC$_{10}$N has a hardness value $\sim 87$ GPa and a relatively low formation energy. Therefore, BC$_{10}$N may be synthesized without high-pressure and high-temperature conditions, for example using low-temperature plasma methods.
	Once synthesized, BC$_{10}$N would become the second known hardest material with a wide range of potential applications in extreme environments.
	
	\section*{ACKNOWLEDGMENTS}
	This research is supported by the U.S. National Science Foundation (NSF) under award OIA-1655280.
	The calculations were performed on the Frontera computing system at the Texas Advanced Computing Center. Frontera is made possible by NSF award OAC-1818253.
	
	\bibliography{bibfile_V1}
	
\end{document}